\documentclass{ws-ijmpd}

\begin{document}

\renewcommand{\baselinestretch}{1.5}
\newcommand{\Prd}{Phys. Rev D}
\newcommand{\Prl}{Phys. Rev. Lett.}
\newcommand{\Pl}{Phys. Lett.}
\newcommand{\Cqg}{Class. Quantum Grav.}
\newcommand{\Sch}{Schwarzschild$\;$}

\newcommand{\half}{\frac 1 2}
\newcommand{\beq}{\begin{equation}}
\newcommand{\eeq}{\end{equation}}
\newcommand{\vare}{\varepsilon}
\newcommand{\ochi}{\overline\chi}
\newcommand{\lmod}{{\cal L}}
\newcommand{\tif}{\tilde{F}}
\newcommand{\lc}{{\cal L}}

\markboth{S. E. Perez Bergliaffa}
{Effective Geometry in Astrophysics}

%
\catchline{}{}{}{}{}
%

\title{EFFECTIVE GEOMETRY IN ASTROPHYSICS}

\author{S. E. PEREZ BERGLIAFFA}

\address{Centro Brasileiro de Pesquisas F\'{\i}sicas,
Rua Dr.\ Xavier Sigaud 150,\\ Urca 22290-180 Rio de Janeiro, RJ -- Brazil
\\
sepb@cbpf.br}

\maketitle

\begin{history}
\received{(Day Month Year)}
\revised{(Day Month Year)}
\end{history}

\begin{abstract}
The effective metric is introduced by means of
two examples (non-linear electromagnetism and hydrodynamics),
along with applications in Astrophysics.
A sketch of the generality of the effect is also given.
\end{abstract}


\section{Introduction}

One of the fundamental building blocks of Einstein's General
Relativity is the fact that every form of matter and energy
couples to gravity. As M. Novello put in a picturesque
way, ``I fall, then I exist''\cite{mario}. If this
universal coupling were broken, gravitation could not have
been identified with the geometry of spacetime, experienced by every form
of matter. This feature
distinguishes gravity from the other forces that we know: no other
interaction has this universal character. In spite of this,
in certain circumstances it would be useful to have
partial geometrization schemes, in which particles
move following geodesics of a metric determined by the
interaction with a background field. Quite unexpectedly, this geometrization
program has been carried out in systems that are very different in
nature: ordinary non-viscous fluids, superfluids, flowing and
non-flowing dielectrics, non-linear electromagnetism in vacuum,
and Bose-Einstein condensates, to name some of them\cite{analog}. As we shall see below,
the underlying feature
shared by these systems is that the behaviour of the fluctuations
around a background solution of the equations of motion
(EOM) is governed by an ``effective
metric''. More precisely, the particles associated to the
perturbations do not follow geodesics of the background spacetime
but of a Lorentzian geometry described by the effective metric,
which depends on the background solution and on the
features of the interaction.

The effective geometry has permitted the simulation of several
configurations of the gravitational
field\footnote{Only some kinematical aspects of General Relativity
can be imitated by this method\cite{visserprl}, but not its
dynamical features (see however Refs.~\refcite{v1} and \refcite{v2}).},
such as wormholes and
closed space-like curves for photons\cite{biwh,ctc}, and warped
spacetimes for phonons\cite{mattuwe}. Particular attention has
been paid to analog black holes\cite{escola}, because these would emit Hawking
radiation exactly as gravitational black holes do, and are
obviously much easier to generate in the laboratory. The fact that
analog black holes emit thermal radiation was shown first by Unruh
in the case of dumb black holes\cite{unruh}, and it is the
prospect of observing this radiation (thus testing the hypothesis
that the thermal emission is independent of the physics at
arbitrarily short wavelengths\cite{unruh}) that motivates the
quest for a realization of analog black holes in the laboratory.

A less explored consequence of the effective metric approach is that
the geometrical tools of
General Relativity can be used to study non-gravitational
systems\cite{viss}. In particular, we shall see below
that in a few instances the idea of effective geometry
has been
applied in Astrophysics. I shall begin by presenting
in Sec. \ref{intro} the basics of the idea of effective
geometry in the example of non-linear
electromagnetism, along with its application to magnetars. In Sec. \ref{hydro} it
will be shown that the notion of effective metric also arises
naturally in hydrodynamics, and we shall see that it can be
fruitfully applied to the problem of accretion onto a black hole.
In an appendix, it will be shown
that the effective geometry is a generic consequence of perturbing
the equations of motion of a given field theory around a fixed
background\cite{v1}. We shall close with a discussion.

\section{The effective metric in non-linear electromagnetism}
\label{intro}

Historically, the first example of the idea of effective metric was presented
by W. Gordon in 1923\cite{gordon}, who showed that
light with wave-vector $k_\mu$ in a moving non-dispersive
medium with slowly varying refractive index $n$ and 4-velocity
$u^\mu$ moves according to
$
g^{\mu\nu}k_\mu k_\nu = 0
$ (see Ref.~\refcite{lp} for details),
where
\beq
g^{\mu\nu}= \eta^{\mu\nu}+ (n^2 - 1) u^\mu u^\nu
\label{gordonm}
\eeq
is the effective metric for this problem.
It must be noted that only photons in the geometric optics
approximation move on geodesics of $g^{\mu\nu}$: the
particles that compose the fluid couple instead to the background
flat metric. After this early contribution, the effective metric occasionally
re-surfaced in the literature\cite{v1}.
A good deal of work has been devoted lately
to non-linear electromagnetism, the case we shall examine next. We shall concentrate in
non-linear
electromagnetic theories in vacuum. Let us start with the action
\beq
S = \int \sqrt{-\gamma}\;
\lc(F) \;d^{4}x ,\protect
\label{N1}
\end{equation}
where $F \equiv F^{\mu\nu}F_{\mu\nu}$,
and $\lc$ is an arbitrary function of $F$. Notice that $\gamma$ is the determinant of the
background metric, which we take in the following to be that of flat
spacetime\footnote{The
same techniques can be applied when the background is curved (see for instance
Ref. \refcite{sns}).}.
Varying this action w.r.t. the potential $A_{\mu}$, related to the field by the
expression $
F_{\mu\nu} = A_{\mu;\nu} - A_{\nu;\mu} = A_{\mu,\nu} - A_{\nu,\mu},
$
we obtain the Euler-Lagrange EOM
\begin{equation}
(\sqrt{-\gamma}\;\lc_{F} F^{\mu\nu})_{;\nu} = 0,
 \protect\label{N2}
\end{equation}
where $\lc_{F}\equiv \frac{\delta \lc}{\delta F}.$
In the particular case of a linear dependence of the Lagrangian with
the invariant $F$ we recover Maxwell's equations of motion.

As mentioned in the Introduction, we want to study
the behaviour of perturbations of these EOM around a fixed background solution. Instead of
using the traditional perturbation method\cite{mattfest}, we shall use
a more elegant method set out by Hadamard\cite{hadamard}, in which the propagation
of high-energy photons is studied by following the evolution of the wave front.
It is assumed that the field is  is continuous through the wave front,
but its first derivative is not.
To be specific,
let $\Sigma$ be the surface of discontinuity defined by the equation
$\Sigma(x^{\mu}) =$ constant.
The discontinuity of a function $J$ through the surface $\Sigma$ will be represented by
$[J]_\Sigma$, and its definition is
$$
[J]_\Sigma \equiv \lim_{\delta\rightarrow 0^+} \left( \left. J\right|_{\Sigma +\delta}
- \left. J \right|_{\Sigma - \delta}\right) .
$$
The discontinuities of the field and its first
derivative are given by
\begin{equation}
[F_{\mu\nu}]_{\Sigma} = 0,   \;\;\;\;\;\;\;\;\;\;\;\;\;\;\;\;\;\;\;\;\;\;
[F_{\mu\nu,\lambda}]_{\Sigma} = f_{\mu\nu}  k_{\lambda},
\protect\label{N8}
\end{equation}
where the vector $k_{\lambda}$ is nothing but the normal to the surface
$\Sigma$, that is, $k_{\lambda} = \Sigma_{,\lambda}.$
Let us apply this technique to the case of a
nonlinear Lagrangian for the electromagnetic field, given by $\lc(F)$. Taking the
discontinuity of the EOM (\ref{N2}), we get
\begin{equation}
\lc_{F} f^{\mu\nu} k_{\nu} + 2 \eta\;\lc_{FF}\;  F^{\mu\nu} k_{\nu} = 0,
\protect\label{N18}
\end{equation}
where $\eta\equiv F^{\alpha\beta}f_{\alpha\beta}. $
Defining $F^*_{\mu\nu} \equiv\half\; \eta_{\mu\nu\alpha\beta}F^{\alpha\beta}$,
where $\eta_{\mu\nu\alpha\beta}$ is the completely antisymmetric
Levi-Civita tensor, the second Maxwell equation is given by ${F^{*\mu\nu}}_{\;\;\;;\nu} = 0$
or equivalently,
\begin{equation}
F_{\mu\nu;\lambda} + F_{\nu\lambda;\mu} + F_{\lambda\mu;\nu} = 0.
\protect\label{N12}
\end{equation}
The discontinuity of this equation yields
\begin{equation}
f_{\mu\nu}  k_{\lambda} + f_{\nu\lambda}  k_{\mu} + f_{\lambda\mu}  k_{\nu} = 0.
\protect\label{N13}
\end{equation}
Multiplying this equation by $k^{\lambda}F^{\mu\nu}$ gives
\begin{equation}
\eta  \;k^2 + F^{\mu\nu} f_{\nu\lambda}  k^{\lambda} k_{\mu} +
F^{\mu\nu} f_{\lambda\mu}  k^{\lambda} k_{\nu} = 0,
\protect\label{N141}
\end{equation}
where $k^2 \equiv k_{\mu} k_{\nu} \gamma^{\mu\nu}. $
Substituting in this equation the term $  f^{\mu\nu} k_{\nu} $ from Eq.(\ref{N18})
yields
\begin{equation}
\eta k^2 - 2 \frac{\lc_{FF}}{\lc_{F}} \eta ( F^{\mu\lambda} k_{\mu} k_{\lambda} -
 F^{\lambda\mu} k_{\mu} k_{\lambda} )=0,
\protect\label{N19}
\end{equation}
which can be written as $g^{\mu\nu}k_\mu k_\nu = 0$, where
\begin{equation}
g^{\mu\nu}= \gamma^{\mu\nu} - 4 \frac{\lc_{FF}}{\lc_{F}} F_{\mu\nu}.
\protect\label{N20}
\end{equation}
We then conclude that the high-energy
photons\footnote{For considerations on the regime in which Hadamard's approach is valid,
see Ref. \refcite{v2}.} of a {\em nonlinear}
theory of electrodynamics with $\lc=\lc(F)$ do not propagate on the
null cones of the background metric but on the null cones of an
{\em effective} metric, generated by the self-interaction of the
electromagnetic field. It is important to realize that all
the quantities in these expressions are evaluated at the background
configuration. In other words, the nonlinear EOM must be solved
for a given source in order to calculate the effective metric
for that background.
Note also that there are two metrics in the problem:
the background metric (which is seen by all every type of matter
but photons), and the effective metric (experienced only by
high-energy photons).

Let us briefly sketch the case in which
the Lagrangian depends also on the invariant $G =
F^{\mu\nu}F^*_{\mu\nu}$. The EOM for a nonlinear
$\lc = \lc(F,G)$ are
\beq
 \left( \lc_F F^{\mu\nu} + \lc_G F^{*\mu\nu}\right)_{;\nu} = 0,
\;\;\;\;\;\;\;\;\;\;\;\;\;\;\;\;F^{*\mu\nu}_{\;\;\;\;;\mu} = 0.
\eeq
Following the procedure employed in the case $\lc=\lc(F)$,
it can be shown that light rays do not follow
geodesics of the background spacetime but of the effective metric\cite{n1}
\beq
g^{\mu\nu}  =   \lc_F\eta^{\mu\nu} - 4 \left[ \left(\lc_{FF}
+\Omega_\pm \lc_{FG} \right) F^\mu_{\;\;\lambda} F^{\lambda\nu}
+  \left(
\lc_{FG} + \Omega_\pm \lc_{GG}\right) F^\mu_{\;\;\lambda}
F^{*\lambda\mu}\right]
\eeq
where
\beq
\Omega_\pm =
\frac{-\Omega_2\pm\sqrt{\Delta}}{2\Omega_1}, \;\;\;\;\;\;\;\;\;\;\;\;\;\;
\Delta =
(\Omega_2)^2 - 4\Omega_1\Omega_3,
\eeq
and $\Omega_i$ ($i=1,2,3$) depend on $\lc_i$ and $\lc_{ij}$, with $i,j
= F,G$.
We see that in general nonlinear electrodynamics displays
bi-refringence. That is, the two polarization states of
the photon propagate in a different way. In some special cases,
there is also bi-metricity (one effective metric for each state)\cite{mattfest}.
Bi-metricity is absent in every theory in which
$\Delta = 0$, as for instance in Born-Infeld\cite{n1}.

It is known that
non-linear EOM for the electromagnetic field arise
in the strong-field regime, when quantum corrections must be taken into account.
Objects like as
magnetars\cite{DT92}, charged black holes, and super-conducting cosmic strings
may have such very intense fields. As an example of
the usefulness of the effective metric in this realm, let us discuss the
corrections to gravitational redshift in magnetars\cite{hermansalim}.
Gravity effects cause the observed energies of the spectral lines of
excited atoms in a compact object to be shifted to lower values by a factor
\begin{equation}
\frac{1}{(1 + z)}  \equiv \left( 1 - \frac{2 G}{c^2} \left[\frac{M}
{R }\right] \right)^{1/2} \; . \label{redshift}
\end{equation}
If, as in the case of magnetars, the object is endowed with magnetic field
larger than the critical value $B_{crit}\approx 10^{14} G$,
corrections to this expression coming
from non-linearities originating in vacuum polarization are to be expected.
These corrections can be calculated using Eq. (\ref{N20}). As shown in Ref.
\refcite{hermansalim}, Eq. (\ref{redshift}) must be replaced by
\begin{equation}
 z + 1 \simeq \frac{ \left(1 - \frac{2 G M} {c^2 R} \right)^{-\frac{1}{2} } }
{ \left[ 1 + \beta B^2 \right]^{\frac{1}{2}} } = \frac{\left(1-0.3\frac{M}{R} \right)^{-\frac{1}{2}} } {\left[1 + 0.19 B^2_{15} \right]^{\frac{1}{2}}}
\label{redshift-C}\;
 \end{equation}
where $M$ is the mass of the star in
units of $M_\odot$,
$R$ its radius in units of 10~km, and $B_{15}$ is the B-field in units of $10^{15}$~G.
We see than that in certain cases the assumed gravitational effects may be in fact
a mixture of gravitation with non-linear electromagnetism. This correction may be
pointing out to inconsistencies in the determination of the quotient mass/ratio, inferred
from the redshift\cite{hermansalim}.

\section{Acoustics in flowing fluids}
\label{hydro}
Another example of the idea of effective metric comes from fluid
mechanics\cite{viss,matarresse}. The equations of motion for
inviscid fluids in a Newtonian background are
the continuity equation,
\beq
\partial_t\rho + \vec\nabla . (\rho\vec v) = 0,
\eeq
and the Euler equation,
\beq
\rho( \partial_t \vec v + (\vec v . \vec\nabla) \vec v) = -\vec p -\rho \vec\nabla\phi
-\rho\vec\nabla\Phi,
\eeq
where $\phi$ is the gravitational potential and $\Phi$ is the potential associated
to other external forces.
Assuming that there is no
vorticity\footnote{The restriction of nonzero vorticity is lifted in
Ref. \refcite{vorti}.} means
that $
\vec v = -\vec\nabla \psi.
$
If the fluid is
barotropic we can write
$
\vec\nabla h = \frac 1 \rho \vec\nabla p.
$
Under these assumptions, the Euler equation reduces to
\beq
-\partial_t\psi + h +\half (\nabla\psi)^2 + \phi + \Phi = 0
\eeq
Next we shall linearize the EOM around a given background using
$
\rho = \rho_0 + \epsilon \rho_1 + O(\epsilon^2),
$
and similar developments for $p$ and $\psi$, where
the background quantities have a 0 subindex.
Keeping terms up to first order in $\epsilon$, we get from the linearized EOM:
\beq
-\partial_t\left(\frac{\partial\rho}{\partial p}\rho_0 (\partial_t \psi_1 +
\vec v_0 . \vec\nabla
\psi_1) \right) + \vec\nabla . \left( \rho_0\vec\nabla \psi_1 -
 \frac{\partial\rho}{\partial p}
\rho_0 \;\vec v_0 (\partial_t \psi_1 +\vec v_0 .\vec\nabla\psi_1) \right) =  0.
\label{messy}
\eeq
This messy equation can be rearranged by
introducing $c_{\rm s}^{-2} = \frac{\partial\rho}{\partial p}$, and the metric
\begin{displaymath}
g_{\mu\nu} =
\frac{\rho_0}{c_{\rm s}}
\left(
\begin{array}{ccc}
-(c_s^2 - v_0^2) & \vdots & -v_0^j \\
\ldots & . & \ldots \\
-v_0^j & \vdots &\delta_{ij}
\end{array} \right).
\end{displaymath}
With these definitions,
Eq.(\ref{messy}) can be written as a wave equation for the scalar field
$\psi_1$, namely $\triangle \psi_1 = 0$,
where $\triangle$ is the
d'Alembertian in curved spacetime, defined in terms of the effective metric
$g_{\mu\nu}$ by
\beq
\triangle\psi_1 =
\frac{1}{\sqrt{-g}}\;\partial_\mu(\sqrt{-g}\;g^{\mu\nu}\partial_\nu\psi_1).
\eeq
As in the case of the non-linear electromagnetic field, the dynamics of the fluctuations
of the scalar field are governed by an effective metric, but note that
fluid particles couple only to the
Minkowski background metric.
Sound waves instead couple only to the {\em effective metric}, which
has
only two degrees of freedom (to be chosen from $\psi_0(x)$,
$\rho_0(x)$, and $c_s(x)$) instead of the
6 degrees of freedom that a solution of Einstein's eqns. may have.
Note also that In Einstein gravity, the metric is determined by the distribution of matter through
Einstein equations.
The effective metric instead is related {\em algebraically} to the distribution of matter.

With a metric to our disposal,
many of the notions of GR (like horizon and ergosphere) can be translated
to this context. For example,
take a closed 2-surface $S$. If $\vec v_0$ is everywhere inward pointing on $S$, and
$\vec v_{0\perp}$ is greater than the local $c_s$, then the sound will be swept
inwards by the flow, and trapped in $S$, which is then an {\em outer-trapped surface}.
The {\em trapped
region} is the region containing outer-trapped surfaces, and the {\em acoustic (future) event
horizon}
is the boundary of the trapped region.

These ideas are relevant in the problem of the
stability of accretion onto a black hole. Moncrief\cite{moncrief} showed
that the perturbations $\psi_1$ of a non-self-gravitating
perfect fluid in stationary potential flow accreting onto a
\Sch black hole are governed by the equation
$\Delta\psi_1 = 0$,
where $\Delta$ is built with the effective metric
$$
\gamma_{\mu\nu} = \frac{n}{h}\left( \frac{c}{v_s}\right) \left[ g_{\mu\nu} +
\left( 1 - \frac{v_s^2}{c^2}\right) u_\mu u_\nu \right].
$$
In this equation, $n$ is the particle number density, $h$ is the enthalpy,
$g_{\mu\nu}$ is the \Sch metric, and the
velocity of sound in the background flow is given by
$$
\left( \frac{v_s}{c}\right)^2 = \left(\frac{\partial p}{\partial \rho}\right)_s.
$$
The causal properties of sound propagation are determined by $\gamma_{\mu\nu}$.
In particular,
the sound cones of $\gamma_{\mu\nu}$ lie \emph{inside} of the light cones defined by
$g_{\mu\nu}$. It is easy to see that
there is a \emph{sonic horizon}, which
in the spherically symmetric case, is located at
$\gamma_{tt}(r_s) = 0$, or
$$
\left( \left. -g_{00} + \frac{v_s^2}{c^2}\right)\right|_{r=r_s} = 1.
$$
Using arguments based on energy-momentum conservation, Moncrief was able to show\cite{moncrief}
that
the norm of $\nabla_\mu\psi_1$ is bounded by the initial value of the
energy, $E_0$.
In particular, no unstable perturbation can exist outside the sonic horizon.

\section{Conclusion}

We have seen two examples in Astrophysics in which
that the concept of effective metric is useful: magnetars
and accretion disks. Other applications of this idea in the same field are
the effect of lensing in charged black holes\cite{vitorio}, the influence of
non-linearities in the velocity of gravitational
waves\cite{mio},  and the stability of stellar coronas and winds\cite{parker}.
The effective metric has been applied even in Cosmology\cite{mattfrw,fischer}
In fact, as shown
in the Appendix, in every problem in which the dynamics is non-linear and
we are interested in the perturbations, the effective metric is bound to
play a role. It is precisely this generality, joined with the power of the
techniques of General Relativity, that makes the effective geometry
a useful tool in several branches of Physics and Astrophysics.

\section*{Acknowledgements}

The author would like to thank FAPERj and ICRA-BR for financial support,
and the members of the Organizing Committee of the IWARA 2003
for their hospitality during the
workshop.

\section*{Appendix}

As shown by Barcel\'o {\em et al.},
the emergence of a curved Lorentzian geometry
is a generic result of the linearization of
a classical field theory around a background.
A sketch of the proof of this statement follows\footnote{See Ref. \refcite{v1}
for details.}.
Let us take a scalar field Lagrangian $\lmod =
{\cal L} (\phi , \nabla\phi)$. Using the expansion
$$
\phi (t,\vec x ) = \phi_0 (t,\vec x ) + \epsilon \phi_1 (t,\vec x ) +
\frac{\epsilon^2}{2} \phi_2(t,\vec x) + O(\epsilon^3),
$$
we can linearize the action for $\lmod$. With the help of
the EOM for the background field, the EOM for the fluctuations are
$$
\partial_\mu \left(\left\{
\frac{\partial^2\lmod}{\partial{(\partial_\mu\phi)\partial(\partial_\nu\phi)}}
\right\}_{\phi_0} \partial_\nu\phi_1 \right) - \left(
\frac{\partial^2\lmod}{\partial\phi\partial\phi} -\partial_\mu
\left\{
\frac{\partial^2\lmod}{\partial(\partial_\mu\phi)\partial\phi}\right\}
\right)_{\phi_0} \phi_1 = 0.
$$
The subindex $0$ means that the quantities are evaluated at
the background.
Defining
$$
\sqrt{g}\;g^{\mu\nu} \equiv \left\{
\frac{\nabla^2\lmod}{\nabla(\nabla_\mu\phi)\nabla(\nabla_\nu\phi)}\right\}_{\phi_0}
,$$ the EOM for the fluctuations takes the form
$$
\left\{ \triangle(g(\phi_0)) - V(\phi_0) \right\} \phi_1 =0,
$$
where $\triangle$ is the D´Alembertian built with $g_{\mu\nu}(\phi_0)$
and the background-dependent potential is given by
$$
V(\phi_0) =
\frac{1}{\sqrt{g}}\left\{\frac{\partial^2\lmod}{\partial\phi\partial\phi}-
\partial_\mu\left\{
\frac{\partial^2\lmod}{\partial(\partial_\mu\phi)\partial\phi}\right\}\right\}.
$$
We obtain the result that \emph{the
kinematics of the fluctuations of the scalar field Lagrangian
$\lmod =
{\cal L} (\phi , \nabla\phi)$
is governed by an effective (curved)
geometry}\footnote{For a generalization of this result to
multicomponent fields, see Ref. \refcite{v2}.}.

\end{document}